\documentclass[a4paper, 12pt]{article}
\usepackage{amssymb} 
\newcommand{\boldalpha}{\mbox{\boldmath$\alpha$}}
\newcommand{\boldnabla}{\mbox{\boldmath$\nabla$}}
\def \vs {\vskip5mm}
\def \ni {\noindent}
\def \be {\begin{equation}}
\def \ee {\end{equation}}
\def \bea {\begin{eqnarray}}
\def \eea {\end{eqnarray}}
\def \xvol {d^3{\rm x}}
\usepackage{graphics}
\begin{document} 
\title{
$\qquad\qquad\qquad\qquad\qquad\qquad\qquad\qquad\qquad $
DIAS-STP-98-13\\$\qquad$\\ 
Localizing the Relativistic Electron} 
\author{A J Bracken$^ {*}$\\
School of Theoretical Physics\\ 
Dublin Institute for Advanced Studies\\
10 Burlington Road\\ Dublin 4, Ireland 
\and 
G F Melloy
\\Department of Mathematics \\
The University of Queensland\\ 
Brisbane 4072, Australia}   
\maketitle 
\begin{abstract}
A causally well-behaved solution of 
the localization problem for the free electron is given, 
with natural
space-time transformation properties,  
in terms of Dirac's position operator ${\bf x}$. 
It is shown that, although {\bf x} 
is not an observable in the usual sense, and 
has no 
positive-energy (generalized) eigenstates, 
the 4-vector density
$(\rho({\bf x},t), {\bf j}({\bf x},t)/c)$ is observable, and can be 
localized arbitrarily precisely about any point in space,
at any instant of time, 
using only positive-energy states. 
A suitable spin operator can be diagonalized
at the same time.     
\end{abstract}


\section{Introduction}

The problem of localization 
in the relativistic quantum mechanics
of a particle with nonzero rest-mass $m$
-- whether arbitrarily precise 
localization is possible, and if so, how it should be described --
is almost as old as relativistic quantum
mechanics itself \cite{Dirac,Schro,Landau,Pryce,Papa,Newton}.  
Despite the efforts of many researchers over the 
intervening years \cite{Wightman,Continue,Bacry,Barut,Heger,Haag},  
the problem continues to attract much discussion 
\cite{Latest}, 
indicating that there is no general acceptance of   
any of the resolutions proposed to date.

A view sometimes expressed is that 
all the difficulties associated with the problem
arise because any attempt to localize a  
particle on a scale small compared to its Compton wavelength
$\lambda_C=\hbar/mc$,
involves an 
uncertainty in energy so large that pair-production becomes 
possible, and a one-particle
description of the physics becomes inconsistent.     
As was pointed 
out by Newton and Wigner \cite{Newton} in their well-known 
paper on the problem, this view ``really denies the possibility
of the measurement of the position'' of a particle.
Some authors have 
considered it appropriate to abandon any
attempt at one-particle localization, and to focus  instead
on local observables associated with 
quantized relativistic fields \cite{Haag}, 
but this approach has evidently 
failed to satisfy 
the many physicists who have continued to investigate the
problem at the one-particle level \cite{Continue,Bacry,Barut,Heger,
Latest}.    
Because the theory of relativity 
is, at its heart,  a theory of relations between
events in space-time, and because it is difficult 
to imagine what can constitute 
an event other than the instantaneous localization of a 
particle, the denial of 
one-particle localizability is hard to accept.

It is important to see that the argument 
regarding pair-production and the Compton wavelength
is inconclusive, because a single 
particle can have an arbitrarily large
energy, and hence an arbitrarily large uncertainty in its energy. 
Therefore, 
the fact that a sufficiently small uncertainty in position 
implies an uncertainty in energy much greater than  
$mc^2$ does not {\em in itself} imply a breakdown of 
the one-particle picture.      
Indeed, Landau and Peierls \cite{Landau}
argued a long time ago that the
minimal uncertainty in position is 
better represented as $\hbar c/{\bar E}$, 
where ${\bar E}$ is a characteristic energy associated
with the measurement, so that arbitrarily precise localization
is not ruled out.     

In what follows, we show that arbitrarily precise localization
of a single free
electron is possible, when described in terms of 
{\em observable} attributes of Dirac's position operator
${\bf x}$ for the electron, in particular 
the
familiar probability density
\be
\rho ({\bf x}) = 
\psi ^{\dagger} ({\bf x})\psi({\bf x})
\label{newdens}
\ee
and probability current density
\be
{\bf j}({\bf x})=
c\psi ^{\dagger} ({\bf x}){\boldalpha}\psi({\bf x})
\label{newcurrent}
\ee
associated with 
Dirac's equation. 
Here ${\bf x}$ is the multiplicative operator acting on 
Dirac 4-spinor functions $\psi({\bf x})$, in the  
Hilbert 
space ${\mathcal H}$ 
with scalar product 
\be
(\psi_1\,,\psi_2)=\int \psi_1^{\dagger}({\bf x})\psi_2 ({\bf
x})\,\xvol\,.
\label{scalarprod}
\ee
It seems the reason 
this simple solution has not been discovered long ago is that  
discussions of the localization problem have been consistently
side-tracked by 
the following 
two related properties of ${\bf x}$,  which appear
to cause insurmountable difficulties:   

\vs\ni
(i) ${\bf x}$ does not leave invariant the subspace 
${\mathcal H}^{(+)}\in{\mathcal H}$ of positive-energy states, 
defined by 
\begin{eqnarray} 
\label{pos} 
H\psi ({\bf x}) =  E({\bf p})\psi ({\bf x})\,,\label {Hpsi}  
\\
\label{HDirac}
H =
 {\boldalpha}\cdot {\bf p} + m \beta\,.    
\end{eqnarray}  
Here 
${\bf p} = -i \hbar \partial/\partial {\bf x}$, 
$E({\bf p}) = c\sqrt{{\bf p}^2 + m^2c^2}$,
and ${\boldalpha}\,,\beta$  are the familiar $4\times 4$ 
Dirac matrices.  The square-root in $E(\bf p)$ is defined using the 
Fourier transform.
    
    \vs\ni
(ii) ${\bf x}$ has no positive-energy (generalized) eigenstates.   
\vs
Because only positive-energy states are allowed to the 
electron, it has been common to interpret (i) 
to mean that ${\bf x}$ cannot represent an observable,   
and to interpret (ii) to mean that, {\it \'a fortiori},
arbitrarily precise localization cannot 
be defined in terms of ${\bf x}$.     

Our solution of the localization problem for the
electron is based on the following
two new results, which we derive below:
\vs
\ni
(I) {\it Despite} (i), {\em the probability density }
$\rho({\bf x})$ 
{\em and associated current density} 
${\bf j}({\bf x})$ 
{\em are observable quantities when $\psi$ is a positive-energy
state, as are 
the `mean ${\bf x}$-coordinates
of the electron' $\langle {\bf x}\rangle
= (\psi\,,{\bf x}\psi)$,
and the `uncertainty in the electron's
${\bf x}$-coordinates,' 
$\Delta_{\bf x}= \langle ({\bf x} - \langle {\bf x}\rangle)^2
\rangle ^{1/2}$.}

\vs\ni
(II) 
{\em 
Despite} (ii), 
{\em sequences of positive-energy states of the electron can be
constructed for which the corresponding sequences of densities
and current densities
approach multiples of} $\delta^3({\bf
x}-{\bf a})$ {\em for 
any 
chosen point} ${\bf a}$ {\em in space, 
and for which the corresponding sequences of mean values}
$\langle {\bf x}\rangle$ {\em and uncertainties}
$\Delta_{\bf x}$ {\em approach } ${\bf a}$ and $0$,
{\em respectively.}
\vs

Positive-energy states of the electron for which the
observable
probability density and current density
(and hence the 
electric
charge density and current density) are arbitrarily sharply
localized,
and for which the observable uncertainty in the particle's
${\bf x}$-coordinates is arbitrarily small,  
surely describe arbitrarily precise
localization of the electron.   
Accordingly, we conclude that 
the electron can indeed be localized arbitrarily sharply 
about  any chosen point, and that localization is properly
described in terms of observable properties of 
the Dirac operator ${\bf x}$ and its
associated densities.  Adding force to this conclusion are the
facts that such a description of localization is causally
well-behaved and has natural space-time 
transformation properties,
as we show below.  

Some comments on (I) and (II) are appropriate at this point. 
In regard to (I), we recall that the desirability of the
existence of a nonnegative
probability density, with an associated current density, was one
of the main motivations for Dirac's development of the equation
which bears his name \cite{Dirac}.  However, because of (i), it
is by no means obvious  that the Dirac densities are observable
for positive-energy states of the electron. 
We present below 
a formal proof of the observability of these quantities.
It is remarkable that this question does  not seem to have 
been addressed
in the past because, when multiplied by the electronic
charge, (\ref{newdens}) and (\ref{newcurrent}) 
are the charge and electric current density of
the particle, and in that form have surely been subject to much
experimental scrutiny.     

The nature and generality of the 
association of 
the observables
of a physical system 
with the self-adjoint operators on a Hilbert space,   
has been much discussed since the earliest days of quantum
mechanics
\cite{Schro2,Dirac,Wigner}. 
In this connection, 
it is notable that $\rho ({\bf x})$ and ${\bf j} ({\bf x})$
are observable,
for while it is true that there is a 
self-adjoint operator ${\bf x}$,  
acting
in the
space ${\cal H}\supset{\cal H}^{(+)}$, 
in terms of which these densities
are defined,
this operator is not an observable in the usual sense,
because of (i), and it has no (generalized) eigenstates at all in
the space of physical states, because of (ii). 
In the case of the relativistic
electron therefore, the relationship
between observables and associated self-adjoint operators is
less direct than in nonrelativistic quantum mechanics. 
In the next section, we suggest the name `indirect observable'
for the operator  ${\bf x}$, because it is not an observable
in the usual sense,
but it nevertheless 
has observable attributes.

The result (II) is very surprising.
Because 
${\bf x}$ has no
positive-energy (generalized) eigenstates,  
we might reasonably expect that
$\Delta_{\bf x}$
cannot be made arbitrarily small if only positive-energy states
are considered.  
Indeed, bearing in mind the argument that is often presented
regarding pair-production,
we might expect that $\Delta_{\bf x}\gtrapprox \lambda_C$ in 
positive-energy states.
It seems that this has,
at least implicitly, 
always been assumed
true, 
but it is not so, and it is 
this
discovery which has enabled us to construct, for the first time,
the localizing
sequences mentioned above. 

The reader may yet feel that the non-existence of an
associated self-adjoint 
position operator having generalized (positive-energy)
eigenstates, 
constitutes  a serious difficulty for any 
localization scheme
defined in terms of ${\bf x}$, a difficulty which 
does not exist in the nonrelativistic case. 
However, we shall show that even in nonrelativistic quantum
mechanics, localization of a particle 
cannot be described
adequately 
in terms of generalized eigenstates of a position operator,
but can be described in terms of localization of 
the probability density and current.
Therefore the
non-existence of generalized eigenstates in the relativistic
case does not represent an  insurmountable 
difficulty for the localization problem, as 
has
commonly been assumed.

The most important 
difference between the cases of the relativistic electron
and the nonrelativistic particle is not the non-existence of a
position operator with generalized eigenstates in the
relativistic 
case.   Rather it is that 
in the nonrelativistic  case, states can be constructed for which the
probability density has compact support.  This is not possible
for the relativistic 
electron, as is well-known \cite{Thaller}; 
all positive-energy wavefunctions, and hence 
probability densities, have `tails' extending to infinity 
in all directions of ${\bf x}$-space.   
These tails typically decay like $\exp (-|{\bf x}|/\lambda_C)$.
It is 
an important consequence of our results that the presence of these
tails does not preclude the
possibility of arbitrarily precise localization using
positive-energy states; in particular
it does not preclude the possibility of constructing
positive-energy states for which $\Delta_{\bf x}$ is arbitrarily
small, and for which the probability outside any chosen compact
region is arbitrarily small
({\it cf.} Fig. 1).      
We think that the presence of these tails
should be accepted as an important and
interesting feature of {\em all} positive-energy states of
the electron, rather than
a defect. Note that we are referring here to `tails' in 
the space of the Dirac coordinate ${\bf x}$, not the
tails in the 
Newton-Wigner coordinate and related concepts of localization
which have been  discussed, in particular
by Hegerfeldt \cite{Heger}.  
The former tails propagate causally in
time,
the latter do not.     

There is another objection which may be raised against 
the use of
the Dirac operator ${\bf x}$ to describe localization of the
electron, namely that each component
${\dot x}_i$,
$i=1\,,2\,,3$, of  
the associated velocity operator
\be
{\dot {\bf x}}= c\boldalpha\,,
\label{velocity}
\ee
has eigenvalues $\pm c$, which seems inappropriate for a
massive particle. 
But,   
like ${\bf x}$, the operator  
${\dot x}_i$
does
not leave ${\cal H}^{(+)}$ invariant.   It has no positive-energy
eigenstates, so that 
its eigenvalues cannot be
observed directly.  
What can be observed is the expectation value
of ${\dot{\bf x}}$,
and as is well known, in any positive-energy state $\psi$,
\be
\langle
{\dot {\bf x}}\rangle
=\langle c^2 {\bf p}/E({\bf p})\rangle\,. 
\label{aveveloc}
\ee
This is appropriate for the propagation of 
a free,  relativistic particle. 

\section{Electron Observables}

Consider a self-adjoint operator $A$ 
on the Hilbert space ${\mathcal H}$ of 4-spinor functions.
Let
$P^{(+)}$ be the self-adjoint projector onto the `physical'
subspace of
positive-energy vectors ${\mathcal H}^{(+)}$,
satisfying
\be
(P^{(+)})^2 = P^{(+)}\,,
\label{projector}
\ee
and suppose that $A$ is regular in the sense that  
\vs\ni
(a) for each 
$k = 1,2\dots$, the operator 
$P^{(+)} A^kP^{(+)}$
is self-adjoint on 
${\mathcal H}^{(+)}$, and so represents an observable for the
electron; and  
\vs\ni
(b)
there exists a common, invariant, dense domain ${\cal D}$ 
of vectors in ${\mathcal H}^{(+)}$ for 
the set of operators 
$P^{(+)} A^kP^{(+)}$,      
$k = 1,2\dots$  

Then if the electron is in a positive-energy state $\psi\in
{\cal D}$,
the expectation value 
\begin{eqnarray}
\langle
P^{(+)} A^k 
P^{(+)}\rangle
 &=&
(P^{(+)}\psi, A^kP^{(+)}\psi)
\nonumber 
\\
\label{expec}
&=& (\psi, A^k\psi)=
\langle
A^k 
\rangle  
\end{eqnarray}
is observable for every $k$. 
But knowledge of all the
expectation values (moments) $\langle A^k\rangle$
is sufficient, except in pathological  cases \cite{Shohat}, 
to determine the distribution of probability over the spectrum
of $A$, consistent with those moments.  For example, if 
$A$ has a discrete  spectrum of eigenvalues 
$a_n\,,n=1,2\dots$ 
corresponding to eigenvectors $\varphi_n\in{\mathcal H}$,
then knowledge of all the moments 
determines the probability  $p_n =
\vert (\phi_n,\psi)\vert ^2$
associated with each eigenvalue $a_n$ of $A$, 
such that    
$
\langle A^k\rangle 
=\sum _n p_n (a_n)^k\,, \,k=1,2,\dots\,
$
 
In this way we see that  it is possible in
principle, 
for each of 
a dense set of positive-energy states of the electron, 
to determine by measurements
a corresponding distribution
of probability over the spectrum  of
the (regular) self-adjoint operator $A$ on ${\mathcal H}$, 
{\it whether or not this operator leaves} ${\mathcal H}^{(+)}$ 
{\it invariant.}    
In the discrete case, it is appropriate to  call 
$p_n$ 
the probability 
`associated with the eigenvalue' 
$a_n$ of $A$ when the electron is in the state $\psi$.
If $A$ \emph{does} leave ${\mathcal H}^{(+)}$ invariant, 
we can go further and call
$p_n$
`the probability that $A$ will be found on 
measurement to 
have the value $a_n$,' because it is then possible that a 
measurement will project 
$\psi$ onto $\phi _n$.

Dirac's coordinate operator ${\bf x}$ is regular in the sense
described. In this case, we have to show firstly that 
$B_{ij\dots k}=
P^{(+)} x_i x_j \dots x_k P^{(+)}$ is self-adjoint, for any
number of terms in the product.
This can be seen by working in momentum space, where
$x_i=i\hbar \partial /\partial p_i$ and $P^{(+)}=(E({\bf
p})+H)/2E({\bf p})$. It is enough to see that the matrix
function $P^{(+)} ({\bf p})$ is hermitian, and that each element
of the matrix is a $C^{(\infty)}$-function, which remains
bounded as $|{\bf p}|\to \infty$.   Then the domain of
self-adjointness of 
$B_{ij\dots k}$ is $P^{(+)}Q\subset Q$, where $Q\subset {\cal
H}$ is the domain of 
self-adjointness of 
$x_i x_j \dots x_k$.   Secondly, we have to find a 
suitable common, invariant, dense
domain for all operators of the form $B_{ij\dots k}$.   This
is provided
by a space of positive-energy states, with both possible spin
values, of the type described in
Section 4.    

It now follows that it is possible
to determine by measurements
the distribution of probability over
the spectrum of ${\bf x}$ when the electron is in (any one of a
dense set of)
positive-energy states, even though ${\bf x}$ does not leave 
${\cal H}^{(+)}$ invariant. In other words,
the probability
density 
(\ref{newdens})
`associated with ${\bf x}$,' is in principle
observable at any time $t$, and by an extension of the argument,
as a function  $\rho({\bf x}, t)$.     
Relativistic-invariance requires that 
this density function can be determined
in any inertial frame, 
and since $\rho$ is one component of the  4-vector
field $(\rho, {\bf j}/c)$, 
where ${\bf j}$ is defined as in (\ref{newcurrent}) at each
instant of time,   
it follows that the function ${\bf j}({\bf x}, t)$
is also observable
in principle.        

Surprising observable 
distributions of probability over the spectrum 
of ${\bf x}$ at any one time, 
that is to say some surprising observable forms for $\rho({\bf x})$,
lie at the heart of our solution below
of the localization problem for the free
electron. These distributions are arbitrarily sharply peaked
about any chosen point in the spectrum of ${\bf x}$, even though
${\bf x}$ has no positive-energy eigenstates.    

It seems to us   unnecessarily restrictive, indeed misleading,
to allow the name `observables'
only for those $A$ which leave ${\mathcal H}^{(+)}$ 
invariant and to remove the others 
from further consideration, 
given that all have observable attributes.    
This is particularly so 
in the case of ${\bf x}$, which has an important
role with an intuitive meaning 
(as the location of charge) when the electron
is coupled to an external electromagnetic field.  
We prefer to follow Dirac \cite{Dirac} and 
call all self-adjoint operators acting on ${\mathcal H}$ 
observables.   Then any observable 
has a (real) spectrum, and a
choice of positive-energy state
of the electron determines an observable distribution of 
probability over that spectrum.
But there is a special subclass of observables,
those which do leave ${\mathcal H}^{(+)}$ invariant.     
These have the further property that they can be diagonalized
on positive-energy states, 
and so can be measured in the traditional sense.
Accordingly, we suggest that an observable be called `direct' if it 
leaves ${\mathcal H}^{(+)}$ invariant, and `indirect' otherwise.     

The reader is  of course free to 
reject this suggestion as a matter of
taste; the main point of the above argument has not been to introduce
the concepts of direct and indirect observables, though we think
that useful, but rather to show
that quantities like the probability density (\ref{newdens}) and
current density (\ref{newcurrent}),   
as well as the expectation values
$\langle {\bf x}\rangle$ and  $\Delta_{\bf x}$,  are observable,
even though ${\bf x}$ is not an observable in the traditional
sense.   

These ideas can obviously be extended 
to the case when an external field
is present, so long as there 
is a well-defined subspace of electron states
in ${\mathcal H}$.    
Whether an observable is direct or indirect may then
depend on the 
field. 
For example, ${\bf p}$ is direct for the 
free electron but indirect
for the electron in a hydrogen atom, 
whereas ${\bf x}$ is indirect in both cases.  
It seems to us a very 
interesting mathematical problem to characterise the possible
{\em observable}
distributions of probability over the spectra of important
indirect observables like ${\bf x}$ 
in such cases.
For example, 
can the probability distribution be arbitrarily sharply peaked
about
any point in the spectrum of ${\bf x}$ for physically-allowed
states of the electron in a
Coulomb field; that is to say, can the electron
be localized arbitrarily sharply in such a field?  
Can the momentum of the
electron in this case be localized arbitrarily sharply?   These
seem to us important questions which should now be tackled.

\section{Localizing a Nonrelativistic Particle}

In nonrelativistic quantum mechanics, say 
for a spinless particle with states described by complex functions
$\chi({\bf q})$, it has become common
to associate localizability of the particle with the existence of a 
generalized
eigenstate 
$
\delta^{(3)}({\bf q} - {\bf a})
$ 
of the position operator ${\bf q}$, for every point
$\bf a$ in space. 
This is inadequate, for two reasons. 

In the first place, 
physically realizable states must be
normalized, and 
no sequence 
$\{\chi_n({\bf q})\}_{n=1}^{\infty}$ 
of normalized  states,
with increasingly sharp localization of ${\bf q}$ about $\bf a$, 
can approach 
$
\delta^{(3)}({\bf q} - {\bf a})
$ 
as $n\to\infty$.  More precisely, no such sequence 
can equal  
the generalized state
$
\delta^{(3)}({\bf q} - {\bf a})\,,
$ 
in the sense of the definition of generalized
functions by sequences \cite{Lighthill}.   
On the other hand, it is clear that a  
sequence of normalized states can be constructed such that
the associated sequence of \emph{densities}
$\{\chi^{*}_n({\bf q})\chi_n({\bf q})\}_{n=1}^{\infty}$
approaches (more precisely,  equals)  
$
\delta^{(3)}({\bf q} - {\bf a})
$,     
with the integral over 
all of ${\bf q}$ - space of each density in the sequence
equal to 1.

In the second place,  it is easily seen that 
it  possible to localize at any chosen time,  about any chosen
point, not only the  
particle's probability density,
but also the associated
current density vector
\begin{equation}
-\frac {i\hbar}{2m} (\chi^{*}({\bf q})
\boldnabla\chi({\bf q}) - 
\chi({\bf q})\boldnabla\chi^{*} ({\bf q}))\,.
\end{equation} 
Consider the sequence of normalized states
defined by
\begin{equation}
\chi_n({\bf q}) =
(n/\sigma\sqrt{\pi})^{3/2}e^{-n^2({\bf q} - 
{\bf a})^2/(2\sigma^2)}e^{im{\bf v}\cdot{\bf q}/{\hbar}}\,,            
\label{NRstates} 
\end{equation}
for $n=1,2,\dots$, 
where $\sigma$, $\bf a$ and $\bf v$ are constants.      
It is easy to check that the associated
sequences of densities and current densities approach (equal)  
$
\delta^{(3)}({\bf q} - {\bf a})
$ and 
$
{\bf v} 
\delta^{(3)}({\bf q} - {\bf a})\,, 
$
respectively.     
{\em This simultaneous
localizability of probability density and current density 
in nonrelativistic quantum mechanics is completely
obscured if localizability is
associated with the generalized eigenstates of} ${\bf q}$.
The densities are not defined even as generalized
functions when  $\chi ({\bf q}) = 
\delta^{(3)}({\bf q} - {\bf a})$.    

The importance of this fundamental point is brought home when
one considers the evolution in time under Schr\"odinger's 
equation for a free particle, 
of a normalized state which is initially 
localized arbitrarily precisely, 
say the state (\ref{NRstates}) for some large value of $n$. 
We find that at time $t\geq 0$, 
\be
\chi_n ^{*}({\bf q},t)\chi_n ({\bf q},t)=
\frac{n\sigma^3}{[\pi(\sigma^4 + n^4\hbar ^2t^2/m^2)]^{3/2}}
e^{-n^2\sigma^2({\bf q}-{\bf v}t)^2/
[\sigma^4 + n^4\hbar^2t^2/m^2]}\,. 
\label{NRtstates}
\ee
This density is 
localized near ${\bf q}={\bf a}$
at $t=0$ and  spreads out as time passes, with a centre that
moves  with constant 
velocity ${\bf v}$.  

In fact, another way to describe the localizing sequence of
states (\ref{NRstates}) is to say that as $n\to\infty$,
$\langle {\bf q}\rangle \to {\bf a}$, $\Delta_{\bf q}\to 0$,
while $\langle {\bf p}/m \rangle \to {\bf v}$. 
In other words, it 
is possible to localize the particle arbitrarily precisely while
at the same time fixing its average velocity at any
chosen value ${\bf v}$;  of course, 
the uncertainty in the
velocity becomes infinitely large as $n\to\infty$.       
Because the velocity 
${\bf p}/m$ is a constant of the motion, 
its mean value remains constant in time, and 
the centre of the wavepacket moves as shown in (\ref{NRtstates}). 

Contrast this with the evolution in time of the generalized
eigenstate $\delta^{(3)} ({\bf q}-{\bf a})$, which gives the
function
\be
G({\bf q},t)=  \left(\frac{m}{2\pi i \hbar t}\right)^{3/2}
e^{im({\bf q}-{\bf a})^2/2\hbar t}\,,\quad t>0 \,.    
\label{Green}
\ee
At any time $t> 0$,  
this is centred on  ${\bf q}= {\bf a}$, and 
is not 
normalizable.    
It   
is clear therefore that
the  generalized eigenstate $\delta^{(3)}({\bf q}-{\bf a})$
does not even approximately describe
an initial state of the particle,
localized sharply near  
${\bf q}={\bf a}$,  
with zero or  nonzero average velocity.   

We conclude that, no matter how useful they may be for other
purposes, 
generalized eigenstates of the position operator are not 
adequate for the  description of localized
states of a particle in nonrelativistic quantum mechanics.

An adequate
notion of localization in the
nonrelativistic case is provided by the introduction of 
localizing sequences of states, as follows.   
Bearing in mind the definition of 
generalized functions by sequences 
\cite{Lighthill}, we call a sequence
$\{\chi_n\}_{n=1}^{\infty}$ of normalized states an 
$({\bf a},
{\bf v})$ - {\em localizing sequence} 
if the associated sequences of 
densities and current densities approach (equal)
$\delta^{(3)}({\bf q} - {\bf a})$
and 
${\bf v}\delta^{(3)}({\bf q} - {\bf a})$, respectively.    
The sequence defined by (\ref{NRstates}) provides an example. 
A particle can be localized arbitrarily precisely about
${\bf q}={\bf a}$, with its probability current localized
along ${\bf v}$ (or equivalently, with its mean velocity equal
to ${\bf v}$), if there exists an 
$({\bf a},
{\bf v})$ - localizing sequence of states.    
If the particle has nonzero spin, localizing sequences
will also carry suitable spin labels.    

Note that the 
sharpness of localization 
is determined by the element of the localizing sequence 
which represents the state of the 
particle  at the chosen time.  Successive elements of the
sequence may be thought of as arising from sharper and sharper  
localizations of the particle at that time.    
(Because the sequence of densities
corresponding to a localizing sequence of states, is equal 
{\em as a whole} to a
delta function \cite{Lighthill}, it is interesting to ask whether or not
it may be sensible to consider
a localizing sequence {\em as a whole}  
to represent an idealized,
exactly localized state of the particle; but we shall
not pursue this
question here.)    

\section{Localizing the Electron} 

The discussion of the preceeding section shows 
clearly that, even in nonrelativistic quantum mechanics,
localization of a particle can  be described adequately only in
terms of arbitrarily sharply localized
probability densities and currents, and not in terms
of generalized eigenstates.   
It follows that, for the localizability 
of the 
relativistic electron 
at any particular instant,  
the important issue is not if 
generalized eigenstates of the Dirac operator 
${\bf x}$ exist -- they do not
-- but whether or not sequences 
$\{\psi_n({\bf x})\}_{n=1}^{\infty}$ 
of {\em positive-energy} normalized states can be found, 
such that the associated sequences of {\em observable}
probability  densities $\rho_n({\bf x})$
and current densities ${\bf j}_n({\bf x})$, 
defined as in (\ref{newdens}) and 
(\ref{newcurrent}),
equal 
$
\delta^{(3)}({\bf x} - {\bf a})
$
and 
$
{\bf v} 
\delta^{(3)}({\bf x} - {\bf a})\,
$,
respectively. 

Remarkably, such $({\bf a}, {\bf v})$ - localizing sequences 
can be found, for every point $\bf a$ in
${\bf x}$ - space, 
and every velocity value $\bf v$ with $\vert {\bf v}\vert<c$,  
despite the fact that ${\bf x}$ has
no positive-energy generalized eigenstates.    
Furthermore, such sequences can
also be chosen to consist of eigenstates
of a suitable spin operator.    
Just as in the nonrelativistic case, 
the sharpness of localization 
is determined by which element of a sequence 
represents the state of the 
electron  at the chosen time.    
As $n\to\infty$, just as in the nonrelativistic case,  
$\langle {\bf x}\rangle \to {\bf a}$, $\Delta_{\bf x}\to 0$,
and $\langle {\bf {\dot x}}\rangle \to {\bf v}$.     

As an example,  consider the sequence defined by 
\begin{eqnarray}
\psi_n({\bf x}) &=& \frac {1}{{(2\pi)}^{3/2}}\int\varphi_n ({\bf p})
e^{i{\bf x}\cdot{\bf p}}\,\, d^3 p\,,
\nonumber
\\
\nonumber
\\
\label{sequence1}
\varphi_n({\bf p}) &=& 
\frac1{n^{3/2}} f(\frac{\bf p}n)u({\bf p})e^{-i{\bf a}\cdot {\bf p}}\,.
\end{eqnarray} 
Here and for the remainder of this section, for simplicity of
presentation we have set $\hbar = c = 1$.  The function
$f({\bf p})$ is a 
a complex-valued, `good' function \cite{Lighthill}, 
that is, a 
$ { C}^{\infty}$-function 
whose  derivatives all 
vanish faster than any negative power of 
$\vert {\bf p}\vert$ 
as 
$\vert {\bf p}\vert\to\infty$.   In addition, we impose 
\begin{equation} 
\label{fconds}
\int \vert f({\bf p})\vert ^2 \,d^3 p =1\,,
\quad
\int \vert f({\bf p})\vert ^2 
\frac {{\bf p}}{\vert {\bf p}\vert}\,d^3 p ={\bf v}\,.   
\end{equation}
The spinor $u({\bf p})$ in (\ref{sequence1})
satisfies 
\begin{equation}
Hu({\bf p})=E({\bf p}) u({\bf p})\,,\quad 
u^{\dagger}({\bf p})u({\bf p}) = 1\,,   
\label{uform}
\end{equation} 
and consequently 
the first of the conditions (\ref{fconds}) ensures that
$\varphi_n$ and hence $\psi_n$ is normalized for every $n$.    
We can  further require that $u$, and hence each $\psi_n$,
is an eigenspinor of some suitable
spin operator 
with eigenvalue $+(1/2)$ or $-(1/2)$, 
which commutes with ${\bf p}$ and $H$,
say for definiteness the third component 
of Pryce's spin operator \cite{Pryce}
\begin{eqnarray} 
{\tilde S}_3({\bf p}) = 
U({\bf p}) (-\frac {i}{2}\alpha _1 \alpha _2)U^{\dagger}({\bf p})\,,
\nonumber
\\
U({\bf p})=(E({\bf p})  I_4+H \beta)/{\cal E({\bf p})}\,,
\end{eqnarray}
where 
${\cal E}({\bf p}) = \sqrt{2E({\bf p})(E({\bf p})+m)}$, and 
$I_4$ is the $4\times 4$ unit matrix

As the Fourier transform of $\psi _n ({\bf x})$ is $\varphi _n({\bf
p})$, so the transform of $\psi _n ^{\dag} ({\bf x})$ is 
$\varphi _n ^{\dag} (-{\bf p})$, and  
the transform of $\rho _n({\bf x})$ (resp. $j_{ni} ({\bf x})$, 
$i=1,2$ or $3$) 
is the convolution
\begin{eqnarray} 
\frac {1}{(2\pi)^{3/2}}
&\int &
\varphi _n ^{\dag} ({\bf q-p})Q\varphi _n ({\bf q}) \,d^3 q
\nonumber
\\
& &(=\frac {e^{-i{\bf a}\cdot {\bf p}}}{(2\pi)^{3/2}}
R_n({\bf p})\,,\rm{say})\,,   
\label{convol}
\end{eqnarray} 
where $Q=I_4$ (resp. $\alpha_i$).   
Noting that the transform of 
$\delta ^{(3)} ({\bf
x}- {\bf a})$ is $e^{-i{\bf a}\cdot {\bf p}} /(2\pi)^{3/2}$, 
we have to show that $\{R_n({\bf p})\}_{n=1}^{\infty}$ equals 
$1$     
(resp. 
$v_i$)  
as a generalized
function. 
From (\ref{sequence1}) and (\ref{convol}) we have 
\begin{eqnarray}
R_n({\bf p})   
&=& \frac1{n^3}
\int
[f(\frac{{\bf q}-{\bf p}}n) u({\bf q}-{\bf p})]^{\dagger}\,
Q f(\frac{\bf q}n) u({\bf q})
\, d^3 q \nonumber
\\
&=& \int
f^{*}({\bf r}-\frac{\bf p}n) 
f({\bf r}) \,
u^{\dagger}(n{\bf r}- {\bf p})
Q
u(n{\bf r})
\,d^3 r\,. 
\label{Rn_form}
\end{eqnarray} 
By Taylor's Theorem, 
\begin{eqnarray}
\label{taylor}
f^{*}({\bf r}-\frac{{\bf p}}{n})
&=&
f^{*}
({\bf r})-
\frac{p_j}{n}
f_j^{*}({\bf r} - \eta \frac{{\bf p}}{n})\,,
\nonumber
\\
u^{\dagger}_a(n{\bf r} - {\bf p})
&=&u^{\dagger}_a(n{\bf r})-p_k u_{ka}^{\dagger}
(n{\bf r} - \theta {\bf p})\,,
\end{eqnarray} 
where $u^{\dagger}_a\,,a=1,2,3,4$, are the components  of $u^{\dagger}$,
$f_j({\bf s})=
\partial f({\bf s})/\partial s_j$, 
$u^{\dagger}_{ka}(
{\bf s}) = 
\partial u^{\dagger}_a({\bf s})/\partial s_k$, 
and $\eta$, $\theta$ are some functions of ${\bf r}$, ${\bf p}$ 
and $n$ (and $a$, in the case of $\theta$)
satisfying  $0\leq\eta\leq 1$, $0\leq\theta\leq 1$.     
In (\ref{taylor}), the summation convention applies to the repeated  
subscripts $j$ and $k$.       
Substituting (\ref{taylor}) in (\ref{Rn_form}), we get
\begin{eqnarray}
R_n({\bf p}) &=&
\int
\vert f({\bf r})\vert^2
u^{\dagger}(n{\bf r})
Q
u(n{\bf r}) \,d^3r
\nonumber
\\
&-& \frac {p_j}{n}\int
f_j^{*}({\bf r}-\frac{{\bf p}}{n})
f({\bf r}) 
u^{\dagger}(n{\bf r})Qu(n{\bf r})\,d^3r     
\nonumber
\\
&-&p_k\int
\vert f({\bf r})\vert^2 
u^{\dagger}_k(n{\bf r} - \theta {\bf p})
Qu(n{\bf r})\,d^3r
\nonumber
\\
&+& \frac {p_jp_k}{n}\int
f_j^{*}({\bf r}-\frac{{\bf p}}{n})
f({\bf r}) 
u^{\dagger}_k(n{\bf r} - \theta {\bf p})
Qu(n{\bf r})\,d^3r
\nonumber
\\
&=& A_n -\frac {p_j}{n}B_{nj}({\bf p}) 
\nonumber
\\
& &- p_k C_{nk}({\bf p})
 + \frac{p_jp_k}{n}D_{njk}({\bf p})\,,\quad\rm{say}\,.
\label{ABCDdef}
\end{eqnarray} 

Making a standard choice of Dirac matrices \cite{Dirac}
with $\beta$ diagonal,
we get for the eigenspinor of 
${\tilde S}_3({\bf s})$ with eigenvalue $+1/2$, 
\begin{equation}
\label{ucomps}
u^{\dagger}({\bf s})=(E({\bf s})+m\,, 0\,,s_3\,,s_1-is_2)/
{\cal E}({\bf s})\,,
\end{equation} 
and it is easily checked that
\begin{eqnarray}
\vert u^{\dagger}_a({\bf s})\vert 
\leq 1\,,
\quad \vert u^{\dagger}_{ka}({\bf s})\vert 
<\frac {2}{m}\,,
\nonumber
\\
\vert u_{ka}^{\dagger}({\bf s})\vert <\frac {2}{\vert{\bf s}\vert}\quad
({\bf s}\ne {\bf 0})\,,
\label{u_inequs}
\end{eqnarray}  
and indeed that the magnitudes of all derivatives of
$u^{\dagger}_{a}$, of all orders,  
are bounded by constants.    
It follows succesively \cite{Lighthill} that 
$\varphi_n ({\bf p})$, $\psi_n({\bf x})$, $\rho_n({\bf x})$, 
${\bf j}_n({\bf x})$ and 
$R_n({\bf p})$  are
good functions for every $n$. 

Using the first and second of (\ref{u_inequs}), 
we see that $ B_{nj}({\bf p})$ and $ D_{njk}({\bf p})$ 
are bounded by constants independent of ${\bf p}$ and $n$.
In $C_{nk}({\bf p})$, 
we note that
if 
$n\vert {\bf r}\vert >2\vert {\bf p}\vert $ $(\,>2\theta 
\vert {\bf p}\vert $), then 
$(n\vert {\bf r}\vert  - \theta \vert {\bf p}\vert )
>\frac {1}{2} n\vert  {\bf r}\vert $, and so 
using the third of (\ref{u_inequs}), 
\begin{equation}
\label{u_inequ1} 
\vert u^{\dagger}_{ka} (n{\bf r} - \theta {\bf p}) \vert
<\frac{2}{\vert n{\bf r} -\theta {\bf p}\vert}
<\frac{4}{n\vert {\bf r}\vert} 
\end{equation} 
whenever $n\vert {\bf r}\vert
 >2\vert {\bf p}\vert $. 
Accordingly, we write
\begin{eqnarray}    
C_{nk}({\bf p}) 
&=&
(\int_{n\vert {\bf r}\vert <2\vert 
{\bf p}\vert } +\int_{n\vert {\bf r}\vert
>2\vert {\bf p}\vert }) 
\nonumber
\\
\nonumber
\\
&\times& \vert f
({\bf r})\vert
^2 
u^{\dagger}_k
(n{\bf r} -\theta {\bf p}) 
Q
u
(n{\bf r})
\,d^3r
\nonumber
\\
\nonumber
\\
&=& C_{nk}^{<}({\bf p}) + C_{nk}^{>}({\bf p})\,,\,\rm{say}\,. 
\end{eqnarray}   
Then, using (\ref{u_inequs}) and (\ref{u_inequ1}), 
\begin{eqnarray}
\vert C_{nk}^{>}({\bf p})\vert 
<
\frac{{\rm const.}}{n} \int_{n\vert {\bf r}\vert >2\vert {\bf p}\vert }  
\frac {\vert f({\bf r})\vert ^2 }{\vert {\bf r}\vert}
\,d^3r
\nonumber
\\
\nonumber
\\
<
\frac {{\rm const.}}{n} 
\int
\frac {\vert f({\bf r})\vert ^2} {\vert {\bf r}\vert}
\,d^3r<\frac{{\rm const.}}{n}\,.   
\label{Bj>}     
\end{eqnarray} 
Furthermore, we have from (\ref{u_inequs}) that 
\begin{eqnarray}
\vert C_{nk}^{<}({\bf p})\vert
&<&{\rm const.} \int_{\vert {\bf r}
\vert <\frac {2\vert {\bf p}\vert }{n}} 
\vert f({\bf r})\vert ^2 \,d^3r 
\nonumber
\\
&<& {\rm const.}
\frac {\vert{\bf p}\vert ^3}{n^3}\,.   
\end{eqnarray} 

Finally, in (\ref{ABCDdef}), 
we see that when $Q=1$ we get 
$A_n=1$ because of the first of (\ref{fconds}); 
and when $Q=\alpha_i$, 
we have 
$
u^{\dagger}(n{\bf r})\alpha_i u(n{\bf r}) =  nr_i/E(n{\bf r})
$, 
so that 
\begin{equation}
\label{Rnlim}
A_n=\int
\vert f({\bf r})\vert ^2 
\frac{  r_i}{\sqrt{\vert {\bf r}\vert ^2 + (m^2/n^2)}}\,d^3r\,,
\end{equation}
which approaches $v_i$ because of the second of (\ref{fconds}).      
From these results for $A_n$, $B_{nj}({\bf p})$, $C_{nk}({\bf p})$ and 
$D_{njk}({\bf p})$,  
it is easy to check that $\{R_n({\bf p})\}_{n=1}^{\infty}$
is a regular sequence \cite{Lighthill} 
which, as a generalized function, 
equals $1$ (resp. $v_i$) as required.     

The choice
$f({\bf p}) = (1/mc{\sqrt \pi})^{3/2} \exp (-{\bf p}^2/2m^2c^2)$ 
in (\ref{sequence1}) leads to a $({\bf 0}, {\bf 0})$-localizing
sequence of positive-energy states, and a corresponding sequence
of
spherically-symmetric probability densities $\rho _n (r)$,
$r=|{\bf x}|/\lambda_C$.   
Figure 1 shows  these increasingly localized densities for 
$n=5$, $7$, $10$,   
as determined numerically with the help of
{\em Mathematica}
\cite{Math}. 
Localization well within the Compton wavelength $(r=1)$ is
evident, even for such small values of  $n$.       
\includegraphics{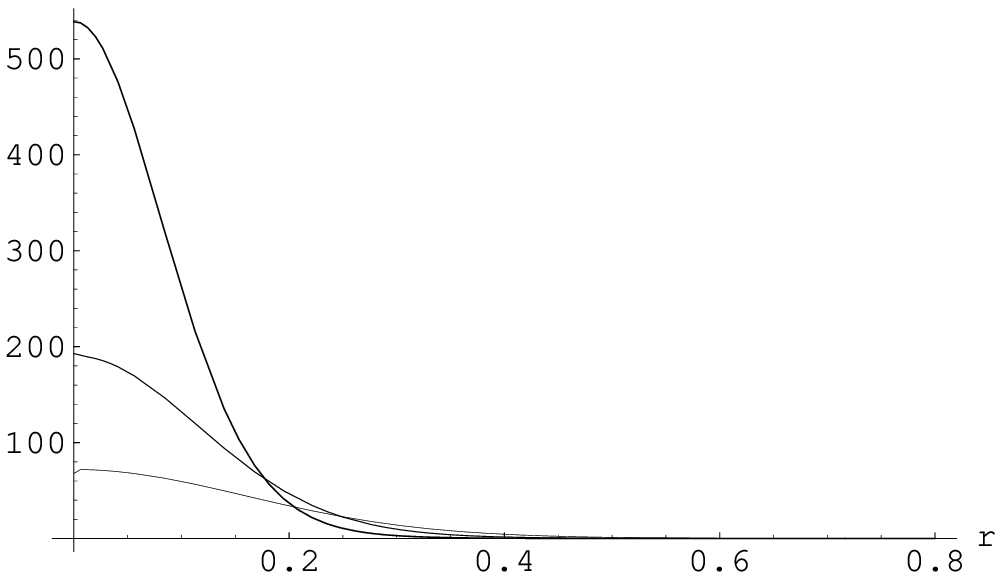}
\begin{center}
{\em Fig. 1}
\end{center}
\section{Space-time Transformations, Causality and
Orthogonality}

Localizing sequences transform naturally under the action of the
(extended) Poincar\'e group.  This is a consequence of the
covariant transformation properties of 
$(ct,{\bf x})$ and $(\rho, {\bf j}/c)$.  

Thus it is easily seen from
(\ref{sequence1})   
that a spatial translation by ${\bf b}$ of
(each element of) an 
$({\bf a},{\bf v})$ - localizing sequence
produces an
$({\bf a} + {\bf b},{\bf v})$ - localizing sequence;
that the parity operation 
produces  a 
$(- {\bf a},- {\bf v})$ - localizing sequence; 
that the operation of time-reversal produces an  
$({\bf a},- {\bf v})$ - localizing sequence;    
and that the result of a rotation by ${\bf R}$ 
produces an 
$({\bf Ra}, {\bf Rv})$ - localizing sequence.   

Lorentz boosts and translations in time require a little more
discussion.    
Consider for example the effect 
of a boost along the 3-axis, 
with $x'_3 =x_3 \cosh\sigma +ct \sinh\sigma$
and $ct'=ct\cosh\sigma +x_3\sinh\sigma$.   
In classical, relativistic physics, if a point-particle  
with charge $e$ is, 
in the original frame, 
at the point ${\bf a}$, with velocity ${\bf v}$,
at (on) the instant $t=0$,   
then in the transformed frame,
the particle is at the point
$(x'_1, x'_2, x'_3)= (a_1, a_2, a_3\cosh\sigma)$,   
with velocity ${\bf v'}= (c/(c\cosh\sigma + v_3\sinh\sigma))
(v_1, v_2, v_3\cosh\sigma  + c\sinh\sigma )$,  
on the hyperplane 
$ct'=x'_3\tanh\sigma$.   
It follows that 
the corresponding charge and current densities are
given at $t=0$ in the first frame by
\be
{\bf j}(0,{\bf x})= {\bf v}\rho (0,{\bf x})=
e{\bf v}\delta^{(3)} ({\bf x}-{\bf a})\,,
\label{class_densities}
\ee
and on the hyperplane 
$ct'=x'_3\tanh\sigma$
in the transformed frame by 
\bea
\nonumber
{\bf j}'((x'_3/c)\tanh\sigma, x'_1, x'_2, x'_3)
= {\bf v}'\rho' ((x'_3/c)\tanh\sigma, x'_1, x'_2, x'_3)=
\\
(v_1, v_2, c\sinh\sigma + v_3\cosh\sigma)
\cosh\sigma\delta
(x_1-a_1)\delta(x_2-a_2)\delta(x_3-a_3\cosh\sigma)\,.
\label{transforms}
\eea
This is mirrored in the effect of the boost 
on an 
$({\bf a},{\bf v})$ - localizing sequence
at (on) the instant
$t=0$.   
From the transformation laws for a 4-vector field, 
we see that 
as the associated sequence of $(\rho, {\bf j}/c)$ values at 
$t=0$ approaches (equals)
$(1, {\bf v}/c)\delta^{(3)}({\bf x}-{\bf a})$, in an obvious
notation, then the sequence of transformed values $(\rho ',
j'_1/c, j'_2 /c, j'_3 /c)$, evaluated  on the hyperplane 
$ct'=x'_3\tanh\sigma$,  
approaches (equals) 
\bea
\nonumber
\lefteqn{(c\cosh\sigma + v_3\sinh\sigma, 
v_1, v_2, c\sinh\sigma + v_3\cosh\sigma)}\\
& &\times 
\frac{\cosh\sigma}{c} \delta
(x_1-a_1)\delta(x_2-a_2)\delta(x_3-a_3\cosh\sigma)\,.
\label{newloc}
\eea

In regard to the evolution of a localizing sequence
in time, we note that, as a consequence of
(\ref{newdens}) and (\ref{newcurrent}), 
and the fact that the eigenvalues
of each $\alpha _i$ are $\pm 1$, 
the Dirac densities satisfy the 
inequality 
$
|{\bf j}({\bf x},t)\cdot 
{\bf n}|\leq  c\rho ({\bf x},t)\,, \label{jdotn}
$
where ${\bf n}$ is any constant unit vector.    
Since the velocity of probability flow is 
${\bf j}({\bf x},t)/\rho({\bf x},t)$,   
this inequality  is a necessary and
sufficient condition for the spread of the probability density $\rho$ 
in every  direction 
to occur 
at speeds no greater than the speed of light.   
Thus the
localization scheme is also guaranteed to behave causally.   

The detailed analysis of how localizing sequences of
positive-energy states evolve in time under Dirac's equation is
an interesting and  nontrivial
problem to which we hope to return in future
work,  
together with an analysis of how 
the spin variable labelling a localizing sequence 
transforms under rotations and Lorentz boosts.

We note finally that localizing sequences
of positive-energy states constructed as above satisfy
a limiting notion of orthogonality:  if 
$\{\psi_n\}_{n=1}^{\infty}$
is an 
$({\bf a},{\bf v})$ - localizing sequence
and $\{\psi_n'\}_{n=1}^{\infty}$ is an  
$({\bf a}',{\bf v}')$ - localizing sequence,   
then it is easily seen from the second of
(\ref{sequence1}), with the help of the 
Riemann-Lebesgue Lemma, that   
\begin{equation}
\label{quasi_orthog}
\lim_{n\to\infty} (\psi_n\,,\psi_n') =0 
\quad\rm{when}\quad
{\bf a}\ne{\bf a}'\,.
\end{equation} 
States from two sequences labelled by different spin eigenvalues
are of course exactly orthogonal to each other.     

\section{Concluding Remarks}

We have shown that the Dirac electron can be localized
arbitrarily sharply about any point in space, at any chosen
instant, when localization is described in terms of 
{\em observable} attributes of Dirac's  operator
${\bf x}$.  

This description of arbitrarily precise (but not exact) 
localization may be compared with the notion of {\em exactly}
localized states 
introduced by Newton and Wigner
\cite{Newton}.   These arise when one  first identifies
an appropriate space
of physical states for a free, relativistic
particle (more precisely, for an elementary system), 
and then seeks to find 
generalized states
satisfying certain conditions which are 
appropriate when the particle is
exactly localized.  
In the case of the electron,
one obtains as a result the  
generalized eigenstates of
the Pryce-Newton-Wigner position operator, which leaves 
the space of physical states invariant.   Because this space 
carries an irreducible, unitary representation 
$R$ (say) 
of the
extended Poincar\'e group, the operator obtained, and the
whole Newton-Wigner
concept of localization, have an invariant group-theoretic
meaning.   This was  emphasized by Bacry \cite{Bacry}, who
showed that the Pryce-Newton-Wigner position operator can be
expressed in terms of the Poincar\'e group generators.  
In this sense, the Newton-Wigner description of localization 
is independent of the {\em realization} of
the representation $R$, and    
there is no particular significance attaching to
the 
realization in terms of Dirac spinors.  

In contrast, out of all 
manifestly covariant realizations of 
$R$ 
with multicomponent wavefunctions $\psi ({\bf x},t)$, 
only 
Dirac's realization 
has an associated nonnegative probability density
$\rho ({\bf x},t)$ 
and associated current density ${\bf j}({\bf x},t)$
satisfying a
conservation equation   
\cite{Barut}.   
These densities play a crucial role in our localization scheme, 
so that Dirac's realization is
distinguished from all others  
when the localization problem 
is considered in the manner which we
have advocated in this paper.   

In our view, localization of the free electron 
can only be described in terms of Dirac's equation and its
associated dynamical variables.
Any 
other realization of $R$ is unitarily equivalent
mathematically to that carried by the positive-energy subspace
${\cal H}^{(+)}\subset{\cal H}$,  but is {\em not equivalent
to it physically}: the
point is that the unitary mapping from the one to
the other is nonlocal in every case \cite{Niederer}.
It is partly  
for this reason that 
investigations of the localization problem 
within the abstract
framework of an irreducible representation of the Poincar\'e group
have not led to the solution that we have  described.    

There is a deeper reason: many investigations of this type have
assumed from the outset 
that the central problem is to find a
self-adjoint operator acting in the space of physical states,
which can
be used to define localization on compact regions,
or exact localization at a point in terms of 
generalized eigenstates.  But in the case of the relativistic
electron,
unlike the case of a nonrelativistic particle, the physical
space of states is a different subspace of ${\cal H}$ for each
different external field. The free particle is just
one special case of this. 
It is for this reason that {\em all} selfadjoint
operators on ${\cal H}$ must be regarded as having 
physical significance;  as we have shown, any one of them 
may have
observable attributes even in cases when it does  not leave the
subspace of physical states invariant, and so is not an
observable in the usual sense.  
Our solution of the localization problem depends on this  more
subtle relationship between operators and observables
which exists for the Dirac electron.  

The noncausal properties of Newton-Wigner localization are
well-known \cite{Wightman}.   A
particle localized in the Newton-Wigner
sense at the origin
at time $t=0$,   
can be found at $t>0$, 
according to the same notion of localization,
outside the sphere of radius $ct$ centred at the origin.
This is unacceptable in a relativistic theory.
It is also known that states exactly localized in the
Newton-Wigner sense do not transform in a simple way with
respect to Lorentz boosts -- in short, because the
Pryce-Newton-Wigner operator is not the three-vector part of a
four-vector. 

It is pleasing therefore that we have been able to show that 
there does exist, at least for the
electron, 
a causally
well-behaved
localization scheme, with natural space-time transformation
properties.
We have emphasized above that 
our analysis has depended critically on the
particular
structure of Dirac's equation, and we 
see no reason to expect that it will extend to other 
particles described by other wave
equations,
excepting  those for the 2- and 4-component neutrinos.   
\vs

AJB thanks L. Bass, P.D. Drummond, R.F. O'Connell and 
L. O'Raifeartaigh for stimulating comments.

\end{document}